\documentclass[aps,prl,twocolumn,nofootinbib,longbibliography,superscriptaddress]{revtex4-1}
\usepackage{hyperref}
\usepackage{amsmath}
\usepackage{amssymb}
\usepackage{amsthm}
\usepackage{graphicx}
\newcommand{\be}{\begin{equation}}
\newcommand{\ee}{\end{equation}}
\newcommand{\ran}{\rangle}
\newcommand{\lan}{\langle}
\newcommand{\bi}{\begin{itemize}}
\newcommand{\ei}{\end{itemize}}
\newcommand{\mO}{\mathcal{O}}

\newcommand{\bfig}{\begin{figure}\begin{center}}
\newcommand{\efig}{\end{center}\end{figure}}

\def\ba#1\ea{\begin{align}#1\end{align}}
\def\bg#1\eg{\begin{gather}#1\end{gather}}

\def\({\left(}
\def\){\right)}
\def\[{\left[}
\def\]{\right]}
\def\<{\langle}
\def\>{\rangle}

\theoremstyle{definition}

\begin{document}

\title{Constraints on symmetry from holography}
\author{Daniel Harlow}
\affiliation{Center for Theoretical Physics\\ Massachusetts Institute of Technology, Cambridge, MA 02139, USA}
\author{Hirosi Ooguri}
\affiliation{Walter Burke Institute for Theoretical Physics\\ California Institute of Technology,  Pasadena, CA 91125, USA}
\affiliation{Kavli Institute for the Physics and Mathematics of the Universe (WPI)\\ University of Tokyo,
   Kashiwa, 277-8583, Japan}
\begin{abstract}
In this letter we show that a set of old conjectures about symmetries in quantum gravity hold within the Anti-de Sitter/Conformal Field Theory (AdS/CFT) correspondence.  These conjectures are that no global symmetries are possible, that internal gauge symmetries must come with dynamical objects that transform in all irreducible representations, and that internal gauge groups must be compact.  These conjectures are not obviously true from a bulk perspective, they are  nontrivial consequences of the non-perturbative consistency of the correspondence.  More details of and background for these arguments are presented in an accompanying paper.  
\end{abstract}

\maketitle
\section{Introduction}
There is an old set of conjectural constraints on symmetries in quantum gravity \cite{Misner:1957mt,Polchinski:2003bq,Banks:2010zn}:
\bi
\item[(1)] Quantum gravity does not allow global symmetries.\label{nosym}
\item[(2)] Quantum gravity requires that there must be dynamical objects transforming in all irreducible representations of any internal gauge symmetry.\label{allcharge}
\item[(3)] Quantum gravity requires that any internal gauge symmetry group is compact.\label{compact}
\ei

None of these conjectures is true as a statement about classical Lagrangians, for example the shift symmetry of a free massless scalar field coupled to Einstein gravity violates conjectures (1) and (3), and the gauge invariance of pure Maxwell theory coupled to Einstein gravity violates conjecture (2). Any argument for these conjectures therefore must rely on properties of non-perturbative quantum gravity.

The ``classic'' arguments for these conjectures are based on black hole physics, but they have various loopholes.  For example until now there was no argument for conjecture (1) which rules out discrete global symmetries such as the $\phi'=-\phi$ symmetry of $\lambda \phi^4$ theory, and all arguments for conjecture (2) require assumptions of some kind about short-distance physics (see \cite{hobig} for more on these arguments).  

The goal of this letter is to use the power of the AdS/CFT correspondence, so far our best-understood theory of quantum gravity, to establish these conjectures, at least within that correspondence.
Along the way we will clarify what we really mean by global symmetry and gauge symmetry, notions which are essential to most of theoretical physics.  In this letter we suppress many details, which are presented in \cite{hobig}.  For simplicity we also discuss only ``internal'' symmetries, which act trivially on the coordinates of spacetime: the analogous statements for spacetime symmetries are again discussed in \cite{hobig}, as are similar statements for higher-form symmetries.

\section{AdS/CFT Review}\label{adssec}
We first briefly review some features of the AdS/CFT correspondence. The basic claim is that any theory of quantum gravity in asymptotically-$AdS_{d+1}$ spacetime, which means a theory where all allowed metric configurations approach the $AdS_{d+1}$ metric
\be
ds^2=-(1+r^2)dt^2+\frac{dr^2}{1+r^2}+r^2 d\Omega_{d-1}^2
\ee
at large $r$, where $d\Omega_{d-1}^2$ is the round metric on $\mathbb{S}^{d-1}$, is nonperturbatively equivalent to a conformal field theory living on the boundary cylinder $\mathbb{R}\times \mathbb{S}^{d-1}$ at $r=\infty$.  The linchpin of the relationship between the two theories is that for any bulk field $\phi$ (suppressing indices), there is a CFT primary operator $\mO$ of scaling dimension $\Delta$ such that \cite{Banks:1998dd,Balasubramanian:1998de,Harlow:2011ke}
\be\label{extdict}
\mO(t,\Omega)=\lim_{r\to\infty}r^{\Delta}\phi(r,t,\Omega)
\ee 
as an operator equation.  For example the boundary stress tensor is the limit of the bulk metric, and the boundary Noether current for a continuous global symmetry is the limit of a bulk gauge field. 

One very important property of the AdS/CFT correspondence is the Ryu-Takayanagi (RT) formula \cite{Ryu:2006bv,Ryu:2006ef,Hubeny:2007xt,Lewkowycz:2013nqa,Faulkner:2013ana,Dong:2016hjy}.  To state this formula, one first needs some geometric definitions. Given a spatial subregion $R$ of the boundary CFT, a codimension-two bulk surface $\gamma_R$ is a Hubeny-Rangamani-Takayanagi (HRT) surface for $R$ if (i) we have $\partial \gamma_R=\partial R$, (ii) the area of $\gamma_R$ is extremal under variations of its location which preserve $\partial\gamma_R$, (iii) there exists an achronal codimension-one bulk surface $H_R$ such that $\partial H_R=\gamma_R\cup R$, and (iv) there is no other surface obeying (i-iii) whose area is less than that of $\gamma_R$.  Generically there will only be one such surface, which we thus will refer to as \textit{the} HRT surface of $R$.  The \textit{entanglement wedge} of $R$ is then defined as $W_R\equiv D(H_R)$, where $D$ denotes domain of dependence and $H_R$ is any bulk surface satisfying the criteria of point (iii).  The (modern version of the) RT formula then says that for any sufficiently semiclassical state $\rho$, the von Neummann entropy of its restriction to the boundary subregion $R$ obeys
\be\label{RT}
S(\rho_R)=\mathrm{Tr}\rho \mathcal{L}_R+S(\rho_{W_R}),
\ee
where $L_R$ is an operator localized on $\gamma_R$ which at leading order in $G$ is given by
\be
\mathcal{L}_R=\frac{\mathrm{Area}(\gamma_R)}{4G}+\ldots
\ee
and $S(\rho_{W_R})$ is the entropy of bulk fields in $W_R$.  

The RT formula has important implications for which bulk degrees of freedom can be represented using only CFT operators in $R$, it was first suspected \cite{Czech:2012bh,Wall:2012uf,Headrick:2014cta} and then proven \cite{Dong:2016eik,Harlow:2016vwg} that in fact \eqref{RT} implies that \textit{any} bulk operator localized in $W_R$ can be represented in the CFT as an operator in $R$: this property of AdS/CFT is called \textit{entanglement wedge reconstruction}. Readers who wish to learn more about this can consult e.g. \cite{Harlow:2018fse}. 

\section{Global symmetry}
As undergraduates we learn that a global symmetry in quantum mechanics is a set of unitary (or possibly antiunitary) operators on the physical Hilbert space which represent the symmetry group and commute with the Hamiltonian.  In quantum field theory however this definition is not satisfactory, for several reasons.  One problem is that some spacetime symmetries, such as Lorentz boosts, do not commute with the Hamiltonian: this is no issue for us in this letter since we discuss only internal symmetries.  More serious is that in quantum field theory the symmetries which are most important are those which respect the local structure of the theory: they must send any operator localized in any spatial region to another operator localized in the same region.  In particular if $U(g)$ are the set of unitary operators representing the symmetry on Hilbert space and $\mO_n(x)$ are a basis for the set of local operators at $x$, then we have
\be\label{Dmap}
U^\dagger(g)\mO_n(x)U(g)=\sum_m D_{nm}(g)\mO_m(x),
\ee
with the matrix $D$ giving an infinite-dimensional representation of the symmetry group $G$.\footnote{In this equation the $\mO_m(x)$ must be smeared against smooth test functions of compact support in order to get genuine operators on the Hilbert space(for internal symmetries this does not affect $D(g)$), and the map among such operators induced by conjugation by $U(g)$ must be continuous in an appropriate sense. These issues are discussed further in \cite{hobig}.}  To make sure we have correctly identified the symmetry group, we require that this representation is \textit{faithful} in the sense that for any $g$ other than the identity, $D(g)$ should also not be the identity.  Finally in quantum field theory global conservation of the symmetry is not enough: we need it also to be locally conserved in the sense that charge cannot be ``teleported'' from one place to another. We can express this mathematically as a requirement that the stress tensor is neutral
\be
U^\dagger(g)T_{\mu\nu}(x)U(g)=T_{\mu\nu}(x),
\ee
which also implies that the symmetry operators $U(g)$ can be freely deformed in correlation functions, at least away from other operators.

One example of a global symmetry is the $\phi'=-\phi$ symmetry of $\lambda \phi^4$ theory which we have already mentioned, and another is $B-L$ symmetry in the standard model of particle physics.  Something which is \textit{not} a global symmetry is the $U(1)$ gauge symmetry of quantum electrodynamics,  for which the map $D$ from \eqref{Dmap} is not faithful since  there are no local operators which are charged (operators which are not gauge-invariant are unphysical and do not count). We will instead interpret this symmetry below as a ``long-range gauge symmetry''.

In quantum field theory we typically expect that continuous global symmetries give rise to Noether currents $J^\mu_a$, where $a$ is a Lie algebra index for the symmetry group.  There is a generalization of this idea to arbitrary global symmetry groups, which we call splittability.  Indeed we say that a global symmetry is \textit{splittable} if, in addition to the operators $U(g)$ we mentioned above, for any spatial region $R$ we have a set of operators $U(g,R)$ which implement the symmetry on operators localized in $R$ and do nothing on operators localized in the complement of $R$ (operators of this type have a history going back to \cite{Doplicher:1982cv,Doplicher:1983if,Buchholz:1985ii}).  In the special case where the symmetry is continuous and has Noether currents, it is always splittable since we can simply take
\be
U(e^{i\epsilon^a T_a},R)\equiv e^{i\epsilon^a\int_R J_a^0}.
\ee  
What these operators do at the edge of $R$ depends on short-distance ambiguities, so we leave this arbitrary.  In fact not all global symmetries in quantum field theory are splittable, and indeed not all continuous global symmetries have Noether currents; in \cite{hobig} we study this phenomenon in some detail.  For this letter however the upshot is that the counterexamples are somewhat pathological, and moreover that in any event they are still ``splittable enough'' for the arguments we make here.  Therefore we will here assume the splittability of all global symmetries without further comment.  

\section{Gauge symmetry}
One of the standard mantras of the AdS/CFT correspondence is that a global symmetry of the boundary CFT is dual to a gauge symmetry of the bulk quantum gravity theory \cite{Witten:1998qj}.  Upon further reflection however this mantra is somewhat puzzling: how can something as concrete as a global symmetry be dual to a mere redundancy of description?  The resolution of this puzzle is that what a boundary global symmetry is really dual to in the bulk is something more refined, which we call a \textit{long-range gauge symmetry}.  

In asymptotically-locally-AdS spacetimes a long-range gauge symmetry with gauge group $G$ is defined by requiring the existence of a set of Wilson lines/loops and asymptotic symmetry operators which obey the following rules.\footnote{One way to develop intuition for these rules is to note in the continuous case that they follow from the usual path-ordered exponential formula for the Wilson lines/loops. In \cite{hobig} they are motivated for general groups using the Hamiltonian formulation of lattice gauge theory.}  First of all for any representation $\alpha$ of $G$ and for any closed curve $C$ in spacetime, we have a Wilson loop operator $W_\alpha(C)$ obeying $W_\alpha(-C)=W_\alpha(C)^\dagger$.  Secondly for any representation $\alpha$ of $G$ and any open curve $C'$ in spacetime with both endpoints on the spatial boundary, we have a Wilson line $W_{\alpha,ij}(C')$, with $ij$ representation indices, obeying $W_{\alpha,ij}(-C')=W_{\alpha,ij}^\dagger(C')$, with ``$\dagger$'' defined to exchange the representation indices.  Moreover $\sum_j W_{\alpha,ij}(-C')W_{\alpha,jk}(C')=\delta_{ik}$.  Wilson lines and Wilson loops are related by a fusing operation where we bring together the endpoints of a Wilson line, take the trace, and then deform away from the boundary.  Thirdly, for any spatial region $R$ of the asymptotic boundary and any element $g$ of $G$, we have a localized asymptotic symmetry operator $U(g,R)$ which acts on any Wilson line from $x$ to $y$, with both $x$ and $y$ in a boundary Cauchy slice containing $R$, as
\begin{align}
U^\dagger(g,R) W_{\alpha} U(g,R)=
\begin{cases}
D_{\alpha}(g)W_{\alpha} D_{\alpha}(g^{-1})& x,y\in R\\
W_{\alpha} D_{\alpha}(g^{-1})& x\in R,y\notin R\\
D_\alpha(g)W_\alpha & x\notin R, y\in R\\
W_\alpha & x,y\notin R
\end{cases},\label{Walg}
\end{align}
where we have suppressed representation indices.  Moreover $U(g,R)$ commutes with any operator localized in the interior of the bulk, and also with its boundary limit provided that it is spacelike-separated from $\partial R$. In particular if $R$ is an entire connected component of the spatial boundary then $U(g,R)$ will commute with the Hamiltonian, and in this case it is called an asymptotic symmetry.  Finally we demand that the ground state is invariant under $U(g,\partial \Sigma)$, where $\partial \Sigma$ denotes the entire spatial boundary, and moreover that the dynamics of the theory allow finite-energy states which are charged under $U(g,\partial\Sigma)$.  A concrete test for this uses the Euclidean path integral in thermal AdS space with inverse temperature $\beta$, with a Wilson line wrapping the thermal circle in the center of the space as well as an asymptotic symmetry operator:
\be
Z_\alpha(g,\beta)\equiv \lan W_{\alpha}(\mathbb{S}^1) U(g,\mathbb{S}^{d-1}\ran_\beta.
\ee   
The test is to see whether or not we have
\be\label{chargetest}
\int dg \chi^*_\alpha(g)Z_\alpha(g,\beta)>0
\ee
for every $\alpha$ and large but finite $\beta$, where $dg$ is the Haar measure on $G$ and $\chi_\alpha(g)\equiv \mathrm{Tr}D_\alpha(g)$ is the character of $g$ in representation $\alpha$.  This object can be interpreted as the thermal trace in a modified Hilbert space with a classical background charge in representation $\alpha$ inserted in the center of the space, and with the integral (or sum if $G$ is discrete) over $g$ implementing a projection onto states transforming in the representation $\alpha$ under the asymptotic symmetry, so \eqref{chargetest} is requiring that such states exist with finite energy.

One example of a theory with a long-range gauge symmetry is Maxwell electrodynamics, which has a $U(1)$ long-range gauge symmetry whether or not there are dynamical charges (dynamical charges are operators which live at bulk endpoints of Wilson lines).  More interesting is quantum chromodynamics (QCD) in $AdS_4$, which has a long-range $SU(3)$ gauge symmetry if its dynamical energy scale $\Lambda_{QCD}$ is small compared to the inverse radius of curvature, but which does not have such a symmetry if $\Lambda_{QCD}$ is large compared to the inverse radius of curvature (since then it is confining and fails the test \eqref{chargetest}).  Many more examples are discussed in \cite{hobig}. 

The notion of long-range gauge symmetry gives a new method for detecting confinement-deconfinement transitions, with \eqref{chargetest} giving the order parameter.  This criterion works even in the presence of dynamical charges transforming in a faithful representation of the gauge group, for example QCD with fundamental quarks, which makes it stronger than the usual ``area-law'' or ``center-symmetry breaking'' tests.  For example in the $\mathbb{Z}_2$ model of \cite{Fradkin:1978dv}, the ``Higgs-confining'' phase is distinguished from the ``free-charge'' phase by the presence of a $\mathbb{Z}_2$ long-range gauge symmetry in the latter, see \cite{hobig} for more discussion of this phase diagram.

\section{Global Symmetries in holography}
We now use these definitions to discuss the symmetry structure of AdS/CFT.  We first argue that any global symmetry in the bulk would lead to a contradiction in the boundary.  The argument uses the entanglement wedge reconstruction reviewed in the second section, and the idea is to show that the existence of a bulk global symmetry would be inconsistent with this.  

\bfig
\includegraphics[height=4cm]{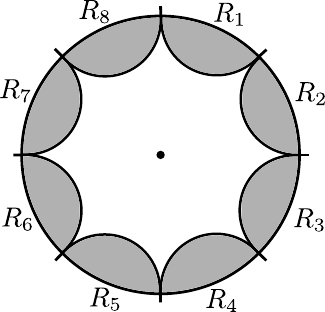}
\caption{A bulk timeslice.  The center of this slice does not lie within the shaded entanglement wedges of the boundary regions $R_i$, so the $U(g,R_i)$ in those regions must commute with any operator there.}\label{regionsfig}
\efig
Indeed say that there were a bulk global symmetry. In AdS/CFT boundary local operators are limits of bulk local operators via \eqref{extdict}, so the local action of a bulk global symmetry implies that its $U(g)$ also give the $U(g)$ of a boundary global symmetry.  By splittability of this boundary global symmetry, we may then split up each $U(g)$ into a product of $U(g,R_i)$ which are localized on some disjoint cover of the boundary spatial slice:
\be
U(g)=U(g,R_1)U(g,R_2)\ldots U(g,R_n) U_{edge}.
\ee 
Here $U_{edge}$ is an operator with support only at the boundaries of the $R_i$ which fixes up the arbitrariness at those boundaries.  The contradiction arises because we can choose all the $R_i$ to be small enough that their associated HRT surfaces $\gamma_R$ do not reach far enough in the bulk for the $U(g,R_i)$ to not commute with an operator in the center of the bulk which is charged under the global symmetry: therefore there can be no localized operators charged under the global symmetry, which is a contradiction (see figure \ref{regionsfig} for an illustration of this).  This contradiction holds even if the operator creating the charged object has large but finite size, such as an operator which creates a black hole of finite energy, since we can always shrink the $R_i$ to pull their entanglement wedges as close to the boundary as we like.\footnote{An important subtlety here is that there are no truly localized operators in a gravitational theory, so we need to define their locations relationally to the boundary using ``gravitational Wilson lines''.  Our argument here can be refined to take this into account, see \cite{hobig} for the details and also for more references on the subject.  Another important subtlety is that the semiclassical picture of the bulk used here is valid only in a ``code subspace'' of the boundary CFT \cite{Almheiri:2014lwa}, this is dealt with also in \cite{hobig} and the contradiction persists.  An interesting connection to a famous theorem in quantum information theory \cite{eastin2009restrictions} is also discussed there.}  

It is instructive to see how this contradiction is avoided for a long-range gauge symmetry in the bulk. In that case, any operator of net charge needs to be attached to the asymptotic boundary by a Wilson line.  This Wilson line will always intersect the entanglement wedge of some one of the $R_i$, so then $U(g,R_i)$ is allowed to detect it.  Indeed if we \textit{assume} the validity of conjecture (2) then it is simple enough to argue that any long-range gauge symmetry in the bulk necessarily implies the existence of a splittable global symmetry in the boundary, with the $U(g,R)$ matching up as expected, and with somewhat less precision one can also argue that the converse holds \cite{hobig}.  We therefore now turn to establishing conjecture (2).
  
\section{Completeness of gauge representations}
Our argument for conjecture (2) is modeled on one presented for the special case $G=U(1)$ in \cite{Harlow:2015lma}.   To establish conjecture (2) in that case, it is enough to show that there is an object of minimal charge; by scattering that object and its CPT conjugate we can create black holes of arbitrary charge.  In fact an analogous statement is true for an arbitrary compact gauge group $G$: given any finite-dimensional faithful representation $\rho$ of $G$, every finite-dimensional irreducible representation appears in the tensor powers of $\rho$ and its conjugate \cite{Levy:2003my}.  Therefore we need only show that when we study the CFT on a spatial $\mathbb{S}^{d-1}$, the bulk asymptotic symmetry $U(g,\mathbb{S}^{d-1})$ acts faithfully on the Hilbert space.  

\bfig
\includegraphics[height=3.5cm]{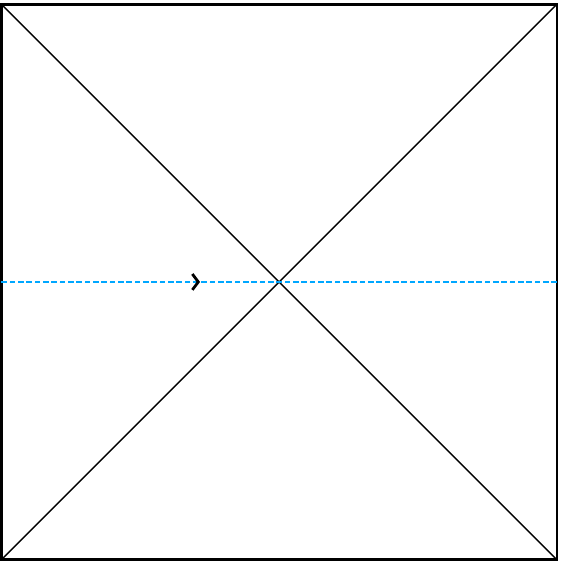}
\caption{A Wilson line threading the AdS-Schwarzschild wormhole}\label{giops2fig}
\efig
The idea is to consider two copies of this system in the thermofield double state.  As explained in \cite{Maldacena:2001kr}, for sufficiently small $\beta$ this state is dual to the maximally-extended AdS-Schwarzschild geometry.  We may then consider a Wilson line $W_\alpha(C)$ on a curve which threads the AdS-Schwarzschild wormhole from one asymptotic boundary to the other, as shown in figure \ref{giops2fig}.  From \eqref{Walg}, 
the algebra of this Wilson line with the asymptotic symmetry operators $U(g,\mathbb{S}^{d-1}_R)$ on the right boundary is 
\be\label{Walg2}
U^\dagger(g,\mathbb{S}^{d-1}_R) W_\alpha(C) U(g,\mathbb{S}^{d-1}_R)=D_\alpha(g)W_\alpha(C),
\ee
where again we have suppressed representation indices.  Now say that the $U(g,\mathbb{S}^{d-1}_R)$ were not faithful: there would then be a $g_0$ not equal to the identity for which $U(g_0,\mathbb{S}^{d-1}_R)=1$. \eqref{Walg2} then would say that $D_\alpha(g_0)=1$ for all $\alpha$.  This however contradicts the Peter-Weyl theorem, which among other things implies that for any group element $g$ other than the identity there will always be some irreducible representation $\alpha$ for which $D_\alpha(g)\neq 1$ \cite{knapp2013lie}.

\section{Compactness}
In the previous section we assumed that the bulk gauge group is compact; in other words we assumed conjecture (3).  In fact this conjecture follows from a simple condition on CFTs.  For simplicity we here discuss only CFTs with a discrete spectrum of primary operators that has no accumulation points, see \cite{hobig} for a discussion of the continuous case.  Our condition is that the CFT in question be \textit{finitely-generated}, which means that there is a finite set of primary operators whose operator product expansion (OPE) recursively generates all of the other primary operators.  Roughly speaking this condition captures the idea that there are a finite number of ``fundamental degrees of freedom'', for example in the theory of a free scalar field in $3+1$ spacetime dimensions all primary operators are polynomials of the scalar and its derivatives.  Now say that a finitely-generated CFT has a global symmetry with symmetry group $G$ (in the bulk this will be a long-range gauge symmetry).  By assumption the $U(g)$ must act faithfully on the set of local operators, and since all of these are generated by our finite set, it must act faithfully already on a finite set of primaries which contains the generating set (the $U(g)$ cannot mix the generating primaries with those of high dimension since they commute with the Hamiltonian).  This possibly larger but still finite set transforms in a finite-dimensional unitary faithful representation $\rho$ of $G$, which we can think of as a homomorphism $\rho:G\to U(N)$ for some $N<\infty$.\footnote{The unitarity of this representation requires an inner product on the set of local operators, which is given by the vacuum two-point function. See appendix C of \cite{hobig} for more details.  One example of a theory where adding an extra operator is necessary is $SO(3)$ current algebra in $1+1$ dimensions, where $J_x$ and $J_y$ already generate all other primaries but we still need to include $J_z$ to fill out the adjoint representation of $SO(3)$.}  The idea is then to observe that the closure of $\rho(G)$ in $U(N)$ is a closed subgroup of a compact Lie group, and therefore by the closed subgroup theorem \cite{lee2001introduction} is itself a compact Lie group.  Moreover by continuity all correlation functions will obey the selection rules of this larger symmetry group: therefore any noncompact global symmetry is part of a larger compact one.  As a simple example, the theory of two compact bosons in $1+1$ dimensions has a $U(1)\times U(1)$ global symmetry rotating the bosons.  This has many one-dimensional noncompact subgroups, each generated by a linear combination of the two charges coefficients whose ratio is irrational.  This theory is finitely generated, and indeed any such subgroup is dense in $U(1)\times U(1)$, which is its closure in $U(2)$ and is of course compact.

\section{Conclusion}
One important issue which these arguments do not touch is approximate global symmetries: our argument for conjecture (1) required assuming an exact global symmetry in the bulk.  In string theory there are many examples of approximate global symmetries, which are violated by Planck-suppressed terms in the low-energy effective action.  It would be very interesting to establish some sort of lower bound on the coefficients of these terms.  Similarly our arguments give no upper bound on the mass of the charged objects required by conjecture (2): some versions of the ``weak gravity conjecture'' of \cite{ArkaniHamed:2006dz} give such a bound, but so far no single version has been convincingly argued for. Such bounds would be very useful for phenomenology, see e.g. \cite{Kamionkowski:1992mf}, and we view the application of AdS/CFT as a tool for establishing them to be a promising avenue for future study.  Finally it clearly is desirable to free ourselves from AdS/CFT and establish conjectures (1-3) for holographic theories on general backgrounds, but this will most likely require a deeper understanding of non-perturbative quantum gravity than is presently available.

\section*{acknowledgments}
{\small 
DH is supported by the US Department of Energy grants DE-SC0018944 and DE-SC0019127, the Simons foundation as a member of the
{\it It from Qubit} collaboration, and the MIT department of physics.
HO is supported in part by
U.S.\ Department of Energy grant DE-SC0011632,
by the World Premier International Research Center Initiative,
MEXT, Japan,
by JSPS Grant-in-Aid for Scientific Research C-26400240,
and by JSPS Grant-in-Aid for Scientific Research on Innovative Areas
15H05895.
}


\bibliography{bibliography}
\end{document}